\documentclass[fleqn,usenatbib]{mnras}

\usepackage{newtxtext,newtxmath,nicefrac}
\usepackage[T1]{fontenc}

\DeclareRobustCommand{\VAN}[3]{#2}
\let\VANthebibliography\thebibliography
\def\thebibliography{\DeclareRobustCommand{\VAN}[3]{##3}\VANthebibliography}

\usepackage{graphicx}	% Including figure files
\usepackage{amsmath}	% Advanced maths commands
\title[TESS and Kepler limb darkening measurements]{Limb darkening measurements from TESS and Kepler light curves of transiting exoplanets}

\author[P. F. L. Maxted]{
Pierre F. L. Maxted$^{1}$\thanks{E-mail: p.maxted@keele.ac.uk} \\
$^{1}$Astrophysics group, Keele University, Staffs, ST5 5BG, UK\\
}

\date{Accepted XXX. Received YYY; in original form ZZZ}

\pubyear{2022}

\begin{document}
\label{firstpage}
\pagerange{\pageref{firstpage}--\pageref{lastpage}}
\maketitle

\begin{abstract}
Inaccurate limb-darkening models can be a significant source of error in the analysis of the light curves for transiting exoplanet and eclipsing binary star systems. 
To test the accuracy of published limb-darkening models, I have compared limb-darkening profiles predicted by stellar atmosphere models to the limb-darkening profiles measured from high-quality light curves of 43 FGK-type stars in transiting exoplanet systems observed by the Kepler and TESS missions. The comparison is done using the parameters $h^{\prime}_1 =  I_{\lambda}(\nicefrac{2}{3})$ and $h^{\prime}_2 = h^{\prime}_1 - I_{\lambda}(\nicefrac{1}{3})$, where $I_{\lambda}(\mu)$ is the specific intensity emitted in the direction $\mu$, the cosine of the angle between the line of sight and
the surface normal vector.  These parameters are straightforward to interpret and insensitive to the details of how they are computed. I find that most (but not all) tabulations of limb-darkening data agree well with the observed values of  $h^{\prime}_1$ and  $h^{\prime}_2$. There is a small but significant offset  $\Delta h^{\prime}_1 \approx 0.006$ compared to the observed values that can be ascribed to the effect of a mean vertical magnetic field strength $\approx 100$\,G that is expected in the photospheres of these inactive solar-type stars but that is not accounted for by typical stellar model atmospheres. The implications of these results for the precision of planetary radii measured by the PLATO mission are discussed briefly.
\end{abstract}


%%%%%%%%%%%%%%%%%%%%%%%%%%%%%%%%%%%%%%%%%%%%%%%%%%

%%%%%%%%%%%%%%%%% BODY OF PAPER %%%%%%%%%%%%%%%%%%

\section{Introduction}

 The variation in the specific intensity emitted from a stellar photosphere with viewing angle is known as centre-to-limb variation (CLV) or limb darkening. Limb-darkening laws typically parametrize the variation in specific intensity at some wavelength $\lambda$, $I_\lambda(\mu)$, as a function of $\mu = \cos(\theta)$, where $\theta$ is the angle between the line of sight and the surface normal vector. 
 For a spherical star, $\mu = \sqrt{1-r^2}$, where $r$ is the radial coordinate on the stellar disc from $r=0$ at the centre to $r=1$ at the limb. Models of eclipsing binary stars and transiting exoplanets typically use limb-darkening laws that assume $I_{\lambda}(1) = 1$. This normalisation is assumed implicitly throughout this paper.

The advent of very high precision photometry for transiting exoplanet systems has led to extensive discussion in the literature of the systematic errors in the parameters for these exoplanet systems that result from inaccuracies and uncertainties in the treatment of limb darkening, e.g. \citet{2013A&A...549A...9C}, \citet{2016MNRAS.457.3573E}, \citet{2013A&A...560A.112M}, \citet{2011MNRAS.418.1165H}, \citet{2008ApJ...686..658S}, \citet{2017AJ....154..111M}, \citet{2017ApJ...845...65N}, \citet{2013MNRAS.435.2152K}, \citet{2022AJ....163..228P}, etc.
One well-established result from such studies is that using a linear limb darkening law, $I_{\lambda}(\mu) = 1- x(1-\mu)$, can lead to significant bias in the parameters derived from the analysis of high quality photometry. For example, \citet{2016MNRAS.457.3573E} found systematic errors in the radius estimates for small planets as large as 3\,per~cent as a result of using a linear limb-darkening law.  The linear limb-darkening law is motivated by a very simple model in which the stellar atmosphere is approximated by a plane-parallel infinite slab with a source function that varies linearly with height.
There are several alternative ways to parametrize limb-darkening that typically add arbitrarily-chosen terms to the linear limb-darkening law to capture the more complex behaviour of real stellar atmospheres. Among the alternative two-parameter laws, the most commonly used in exoplanet studies is the
quadratic limb-darkening law \citep{1950HarCi.454....1K} --
\begin{equation}
I_{\lambda}(\mu) = 1- u(1-\mu) - v(1-\mu)^2.
\end{equation}
This limb-darkening law has the advantage of being relatively simple and
well-understood in terms of the correlations between the coefficients
\citep{2008MNRAS.390..281P, 2011ApJ...730...50K, 2011MNRAS.418.1165H} and how
to sample the parameter space to achieve a non-informative prior
\citep{2013MNRAS.435.2152K}, but it fails to match optical high-precision light
curves of transiting exoplanet systems \citep{2007ApJ...655..564K}.

Among the limb-darkening laws with two coefficients, the power-2 limb-darkening law  \citep{1997A&A...327..199H} has been recommended by \citet{2017AJ....154..111M} as they find that it outperforms other two-coefficient laws adopted in the exoplanet literature in most cases,
particularly for cool stars. The form of this limb-darkening law is
\begin{equation}
I_{\lambda}(\mu) = 1-c\left(1-\mu^{\alpha}\right).
\end{equation}
In \citet{2018A+A...616A..39M} I used this limb-darkening law to analyse high-quality light curves for 16 solar-type stars with transiting hot-Jupiter companions. I found that the parameters $c$ and $\alpha$ are strongly correlated with one another so, to compare these results to the limb-darkening profiles from stellar atmosphere models, I introduced the parameters
\begin{equation}
\begin{array}{ll}
h_1 = & I_{\lambda}(\nicefrac{1}{2})= 1-c\left(1-2^{-\alpha}\right),\\
h_2 = & h_1 - I_{\lambda}(0) = c2^{-\alpha}.
\end{array}
\end{equation} 
These parameters were found to be uncorrelated and so could be use to define useful priors for a Bayesian analysis of a light curve for an eclipsing binary star or transiting exoplanet for solar-type stars using the power-2 limb-darkening law. \citet{2019RNAAS...3..117S} note that the range of valid $h_1$ and $h_2$ values given in \citet{2018A+A...616A..39M} is incorrect. They provide equations to calculate the transformed parameters $0<q_1<1$ and $0<q_2<1$ that span the full range of valid  $h_1$ and $h_2$ values.

\begin{figure}
\begin{center}
\includegraphics[width=0.8\columnwidth]
{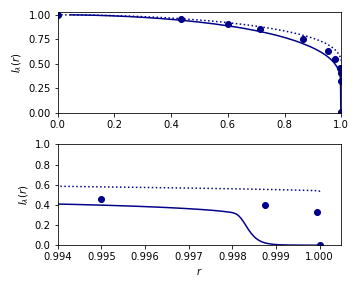}
\end{center}
    \caption{Limb-darkening profiles in the {\it Kepler} band for a star with $T_{\rm eff} = 6000$\,K, $\log g = 4.0$ and solar metallicity from \citet{2013A+A...556A..86N} using spherical-symmetric (solid line) and plane-parallel (dotted line) model stellar atmospheres. Points show the limb-darkening profile in the {\it Kepler} band computed using a 3-d radiative hydrodynamical model stellar atmosphere with T$_{\rm eff} = 6015$\,K at the same surface gravity and metallicity \citep{2015A+A...573A..90M, 2018A+A...616A..39M}.  }
    \label{fig:CLVPlot}
\end{figure}

The definition of $h_2$ causes problems if we want to apply the results from \citet{2018A+A...616A..39M} to other limb-darkening laws, or to test the accuracy of limb-darkening laws computed with model stellar atmospheres. One problem is the definition of the radius for limb-darkening profiles computed with spherically-symmetric model atmospheres. One such profile is shown in Fig.~\ref{fig:CLVPlot}. These models can reproduce the smooth, sharp drop in flux near the limb of the star that is expected for solar-type stars based on observations of the solar limb-darkening profile. 
This drop in flux near the limb is not seen in 1-d plane-parallel model atmospheres. 3-d radiative hydrodynamical models are computed on a grid of finite size, so it is difficult to compute the drop in flux at large values of $\theta$ using these simulations. 
The detailed shape of the drop in flux near the limb has a negligible impact on the observed light curves for solar-type stars, but it does make the definition of $h_2$ ambiguous. This complicates the interpretation of $h_2$ values inferred from light curve analysis. 
 
% CLVPlot

In this study, I have used the Claret 4-parameter limb-darkening law \citep{2000A+A...363.1081C} to analyse high-quality light curves for sample of transiting hot-Jupiter systems. This limb-darkening law uses four coefficients to capture the detailed shape of the limb-darkening
profile using the following equation --
\begin{equation}
\label{eqn:claret}
I_{\lambda}(\mu) = 1 - \sum_{j=1}^{4}a_j(1-\mu^{j/2}).
\end{equation}
I introduce two new parameters, $h^{\prime}_1$ and $h^{\prime}_2$, that can be unambiguously computed for any limb-darkening profile. In Section~\ref{sec:simulations} I use simulations to show that the values of $h^{\prime}_1$ and $h^{\prime}_2$ measured from light curves are directly related to the true limb-darkening profile of the star, and so can be used directly to compare the limb-darkening profiles recovered from the observed light curves to limb-darkening calculated with stellar model atmospheres. Section~\ref{sec:analysis} describes the methods used to analyse the observed light curves for a sample of transiting hot-Jupiter systems, and present the results of this analysis. Section~\ref{sec:discussion} compares these results to previous studies, and to the predictions from a selection of stellar atmosphere models. Section~\ref{sec:conclusions} contains my conclusions and recommendations for how to use these results to constrain limb-darkening in the analysis of light curves for eclipsing binary stars and transiting exoplanet systems. 

\section{Light curve simulations} \label{sec:simulations}
To better understand the constraints on stellar limb darkening provided by the light curves of transiting exoplanets, I used simulations based on the solar limb-darkening profile computed by \citet{2022arXiv220606641K}. This semi-empirical limb-darkening profile uses a combination of observations and model spectra to produce a realistic set of intensity spectra as a function of wavelength and viewing angle. The spectra are computed at the same values of $\mu$ (20 values from 0.01 to 1) for which observed values of the centre-to-limb variation of the Sun are provided by \citet{1994SoPh..153...91N}, sampled at 7000 wavelength values from 339\,nm to 1087\,nm. I used these spectra to compute the limb-darkening profile of a Sun-like star in the {\it Kepler} bandpass\footnote{\url{https://nexsci.caltech.edu/workshop/2012/keplergo/kepler_response_hires1.txt}} by numerical integration of the intensity spectra over the instrument response function, including a factor $hc/\lambda$ to account for the fact that the {\it Kepler} instrument used photon-counting detectors. The 20 values of  $I_{\rm Kp}(\mu)$ obtained were then scaled by a constant so that $ I_{\rm Kp}(1) = 1$.

 I used {\tt batman} version 2.4.7 \citep{2015PASP..127.1161K} to simulate light curves of transiting exoplanets assuming a limb-darkening profile of the form 
 \begin{equation}
 \label{eqn:poly}
 I_{\rm Kp}(\mu) = 1 - \sum_{i=1}^{6} c_i(1-\mu)^i.
 \end{equation}
%For use in {\tt batman}, the constant $I_0$ is required to have a value such that the total intensity integrated over the stellar disc is 1, i.e.   %
%\[ I_0 = \pi\left(1-c_1/3-c_2/6-c_3/10-c_4/15-c_5/21-c_6/28\right). \]
 The coefficients $c_1, \dots, c_6$ were computed using a least-squares fit to the 20 values of $I_{\rm Kp}(\mu)$ described above. The standard deviation of the residuals for this least-squares fit is 13\,ppm.

 To check the accuracy of these simulated light curves, I computed the light curve due to the transit of a planet with a radius ratio $k=R_p/R_{\star} = 0.08$ and a stellar radius $R_{\rm star}/a = 0.15$ assuming a transit impact parameter\footnote{$b=a\cos(i)/R_{\star}$ for a planet with an orbital inclination $i$ in a circular orbit with semi-major axis $a$ around a star of radius $R_{\star}$.} $b=0.4$ using {\tt batman}, and compared this to a light curve computed using {\tt ellc} \citep{2016A&A...591A.111M} using  the ``very fine'' numerical grid option. The results agree to better than 4\,ppm at all phases, and to better than 1\,ppm outside the ingress and egress of the transit.

 I then used {\tt batman} to simulate light curves for a range of $b$ values from $b=0$ to $b=0.8$ for the same values of $k$ and $R_{\star}/a$ noted above. These simulated light curves were sampled at 1000 points uniformly distributed across the duration of the transit. For each value of $b$, I generated 1000 light curves including Gaussian random noise with a standard deviation of 100\,ppm per observation. 
 This is similar to the signal-to-noise in {\it Kepler} light curves of moderately bright stars with transiting hot Jupiters, e.g.  Kepler-5. I then did a least-squares fit to these simulated light curves using Claret's 4-parameter law to model the limb darkening. The free parameters in these least-squares fits were $k$, $b$, $R_{\star}/a$, and the four limb-darkening coefficients $a_1,\dots, a_4$. These parameters are strongly correlated with one another, which can be problematic for many least-squares optimisation algorithms. To quantify these correlations I used the affine-invariant Markov-chain Monte-Carlo sampler {\tt emcee} \citep{2013PASP..125..306F, GoodmanWeare2010} to sample the posterior probability distribution of these parameters for the least-squares fit to one simulated light curve. I then used principal component analysis (PCA) as implemented in {\tt scikit-learn} \citep{scikit-learn} to find a linear transformation between the free parameters of the model and seven uncorrelated variables $p_1\, \dots, p_7$. To find the best fit to each simulated light curve I used the Nelder-Meade algorithm implemented in {\tt scipy} \citep{2020SciPy-NMeth} with the simplex defined in the transformed parameter space $p_1\, \dots, p_7$. For every trial set  of limb-darkening coefficients, $a_1, \dots, a_4$, the limb-darkening profile was computed at 100 uniformly-distributed values of $\mu$ from 0.01 to 1. Solutions where these coefficients do not correspond to a physically realistic limb-darkening profile with $0 < I(\mu) <1 $ and $\frac{dI(\mu)}{d\mu} > 0$ for all values of $\mu$ were rejected.

% RecoveredProfile
\begin{figure}
	\includegraphics[width=\columnwidth]{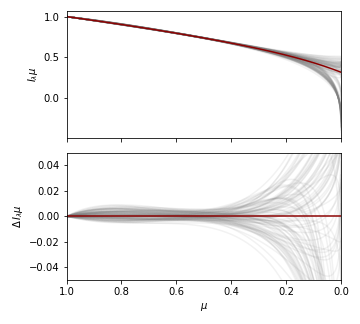}
    \caption{Limb-darkening profiles for a solar-type star recovered from 100 simulated light curves for a transiting hot Jupiter system with the following parameters: $k=0.08$, $R_{\rm star}/a = 0.15$, $b=0.4$. The red line in the upper panel shows the limb-darkening profile used to generate the simulated light curve. The lower panel shows the deviation from this assumed limb-darkening profile.}
    \label{fig:RecoveredProfile}
\end{figure}

 The limb-darkening profile recovered from one simulated light curve with $b=0.4$ is shown in Fig.~\ref{fig:RecoveredProfile}. The limb-darkening profiles for other values of $b\loa 0.6$ are qualitatively similar. 
 It is clear from this figure that the simulated light curve contains very little information about the limb-darkening profile of the star for $\mu \loa 0.2$, i.e. near the limb of the star. 
 This may seem to be at odds with the results from  \citet{2018A+A...616A..39M} where values of $h_2 = I_{\rm Kp}(\nicefrac{1}{2}) - I_{\rm Kp}(0)$ are quoted with a typical accuracy of about $\pm 0.05$. 
 However, those results were based on least-squares fits to {\it Kepler} light curves of transiting hot Jupiters assuming a power-2 limb-darkening law. The power-2 limb-darkening law has only two parameters. The implication of Fig.~\ref{fig:RecoveredProfile} is that the parameter $h_2$ is determined by the limb darkening profile at $\mu \goa 0.2$, even though it is defined in terms of $I_{\rm Kp}(0)$. So, as well as being ambiguously defined, the definition of $h_2$ is also misleading, in that it does not measure what claims to measure.  The value of $h_2$ will also be subject to systematic error if the true limb-darkening profile does not closely match the assumed limb-darkening law near the limb of the star.

% AccuracyTest
\begin{figure}
\includegraphics[width=\columnwidth]{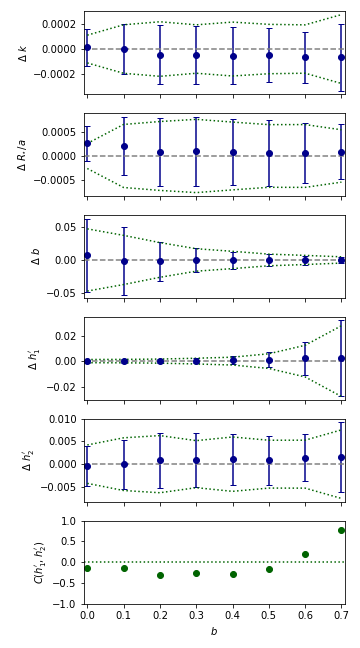}
\caption{Transit-model and limb-darkening parameters recovered from simulated light curves for a transiting hot Jupiter system with the following parameters: $k=0.08$, $R_{\rm star}/a = 0.15$. Green dashed lines in each panel show the standard error on each parameter estimated from the PPD sampled with {\tt emcee} for one simulation at each value of $b$. Blue points with error bars show the mean and standard error of the results from 1000 simulations. The points in the bottom panel show the correlation coefficients for the parameters $h^{\prime}_1 = I_{\rm Kp}(2/3)$ and $h^{\prime}_2 = h^{\prime}_1 - I_{\rm Kp}(1/3)$ evaluated from the PPD sampled with {\tt emcee}  for one simulation at each value of $b$.}
    \label{fig:AccuracyTest}
\end{figure}

\begin{figure}
	\includegraphics[width=\columnwidth]{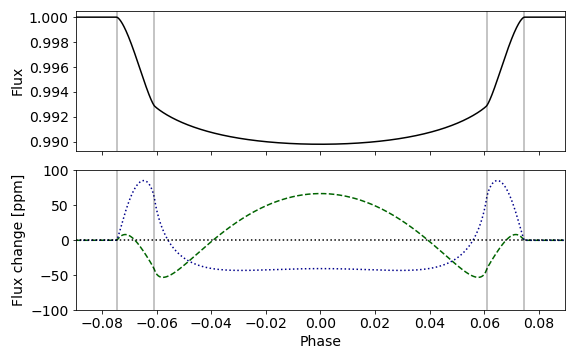}
    \caption{Upper panel: simulated light curve for the following parameters:
  $R_{\star}/a=0.125$, $k=0.093$, $b=0.3$, $h^{\prime}_1 = 0.835$, $h_2^{\prime} =
  0.186$. Lower panel: change in flux due to a change of +0.01 in the value of  $h^{\prime}_1$ (dashed line) or $h_2^{\prime}$ (dotted line). Vertical lines indicate the contact points of the transit.}
    \label{fig:h1ph2plc}
\end{figure}

Based on these results, I have decided to use the parameters 
\begin{equation}
\begin{array}{ll}
h^{\prime}_1 = & I_{\lambda}(\nicefrac{2}{3})\\
h^{\prime}_2 = & h^{\prime}_1 - I_{\lambda}(\nicefrac{1}{3})
\end{array}
\end{equation} 
to compare the limb-darkening measured from transit light curves to limb-darkening profiles computed from models.  The choice of $\mu=\nicefrac{1}{3}$ and $\mu=\nicefrac{2}{3}$ for these definitions is arbitrary but these values do correspond to points where the recovered limb-darkening profiles are quite well determined, and numerical experiments show that the correlation between $h^{\prime}_1$ and $h^{\prime}_2$ is generally quite low for these values of $\mu$.
The values of $h^{\prime}_1$ and $h^{\prime}_2$ recovered from the simulated light curves are shown as a function of $b$ in Fig.~\ref{fig:AccuracyTest}. These results show that the values of $h^{\prime}_1$ and $h^{\prime}_2$ obtained by fitting a transit light curve using Claret's 4-parameter limb-darkening law are accurate, i.e. the bias in the mean value (offset from the true value) is small compared to the uncertainty on these values. Fig.~\ref{fig:AccuracyTest} also shows that $h^{\prime}_1$ and $h^{\prime}_2$ are not strongly correlated for transit impact parameter values $b \loa 0.65$. For  $b \goa 0.65$, $h^{\prime}_1$ and $h^{\prime}_2$ are strongly correlated because the light curve does not contain enough information to determine these parameters independently. One further conclusion we can take from Fig.~\ref{fig:AccuracyTest} is that the standard error estimates on $h^{\prime}_1$ and $h^{\prime}_2$ based in the PPD sampled with {\tt emcee} are accurate for a light curve with uncorrelated Gaussian noise. Fig.~\ref{fig:h1ph2plc} shows how a small change in the value of $h^{\prime}_1$ or $h^{\prime}_2$ changes the shape of the light curve for a typical transiting hot-Jupiter system.

 I used simulations similar to those described above to verify that the values of $h^{\prime}_1$ and $h^{\prime}_2$ obtained by fitting a transit light curve are insensitive to ``third light'' (contamination of the light curve by an unresolved companion star or other light source) provided the contamination is less than about 5\,per~cent. Similarly, I found that the results are also not badly affected by assuming a circular orbit for systems where the true orbital eccentricity is small ($e\loa 0.1$).
  
% h1ph2plc figure

\section{Analysis} \label{sec:analysis}

% Star properties table
\begin{table}
	\centering
	\caption{Effective temperature (T$_{\rm eff}$), surface gravity ($\log g$ in cgs units) and metallicity ([Fe/H]) for the stars analysed in this study.}
	\label{tab:stars}
 \addtolength\tabcolsep{-0.25pt} % Slight squeeze on column widths
\begin{tabular}{@{}lrrrr} 
\hline		
Star & 
  \multicolumn{1}{l}{KIC} &
  \multicolumn{1}{c}{T$_{\rm eff}$ [K] } &
  \multicolumn{1}{c}{$\log g$ } & 
  \multicolumn{1}{c}{[Fe/H]} \\
		\hline
HAT-P-7      &     10666592 & $6575 \pm   34$ & $4.08 \pm 0.01$ & $+0.28 \pm 0.02$ \\
Kepler-4     &     11853905 & $5885 \pm   33$ & $4.15 \pm 0.01$ & $+0.19 \pm 0.03$ \\
Kepler-5     &      8191672 & $6297 \pm   60$ & $4.17 \pm 0.02$ & $+0.04 \pm 0.06$ \\
Kepler-6     &     10874614 & $5647 \pm   50$ & $4.28 \pm 0.02$ & $+0.34 \pm 0.05$ \\
Kepler-7     &      5780885 & $6213 \pm   41$ & $4.03 \pm 0.02$ & $+0.19 \pm 0.03$ \\
Kepler-8     &      6922244 & $6181 \pm   60$ & $4.18 \pm 0.02$ & $-0.08 \pm 0.04$ \\
Kepler-12    &     11804465 & $5926 \pm   60$ & $4.14 \pm 0.02$ & $+0.03 \pm 0.04$ \\
Kepler-14    &     10264660 & $6500 \pm   60$ & $4.22 \pm 0.10$ & $+0.05 \pm 0.04$ \\
Kepler-15    &     11359879 & $5662 \pm   60$ & $4.19 \pm 0.10$ & $+0.27 \pm 0.04$ \\
Kepler-17    &     10619192 & $5667 \pm   63$ & $4.48 \pm 0.03$ & $+0.18 \pm 0.05$ \\
Kepler-40    &     10418224 & $6296 \pm  104$ & $4.04 \pm 0.04$ & $+0.01 \pm 0.07$ \\
Kepler-41    &      9410930 & $5766 \pm  118$ & $4.31 \pm 0.05$ & $+0.21 \pm 0.09$ \\
Kepler-43    &      9818381 & $6150 \pm   90$ & $4.35 \pm 0.04$ & $+0.41 \pm 0.07$ \\
Kepler-44    &      9305831 & $5757 \pm  134$ & $4.36 \pm 0.06$ & $+0.26 \pm 0.10$ \\
Kepler-45    &      5794240 & $3820 \pm   90$ & $4.53 \pm 0.08$ & $+0.20 \pm 0.10$ \\
Kepler-74    &      6046540 & $6056 \pm   62$ & $4.37 \pm 0.03$ & $+0.34 \pm 0.05$ \\
Kepler-77    &      8359498 & $5595 \pm   60$ & $4.47 \pm 0.03$ & $+0.37 \pm 0.04$ \\
Kepler-412   &      7877496 & $5875 \pm   49$ & $4.29 \pm 0.02$ & $+0.27 \pm 0.04$ \\
Kepler-422   &      9631995 & $5891 \pm   60$ & $4.32 \pm 0.03$ & $+0.21 \pm 0.04$ \\
Kepler-423   &      9651668 & $5790 \pm   80$ & $4.51 \pm 0.04$ & $+0.26 \pm 0.05$ \\
Kepler-425   &      5357901 & $5170 \pm   70$ & $4.55 \pm 0.04$ & $+0.24 \pm 0.11$ \\
Kepler-426   &     11502867 & $5535 \pm   60$ & $4.47 \pm 0.03$ & $-0.17 \pm 0.04$ \\
Kepler-427   &      7950644 & $5800 \pm   70$ & $4.24 \pm 0.03$ & $-0.19 \pm 0.07$ \\
Kepler-428   &      5358624 & $5150 \pm  100$ & $4.64 \pm 0.05$ & $+0.09 \pm 0.17$ \\
Kepler-433   &      5728139 & $6360 \pm  140$ & $4.09 \pm 0.05$ & $-0.01 \pm 0.19$ \\
Kepler-435   &      7529266 & $6388 \pm   45$ & $4.23 \pm 0.07$ & $+0.06 \pm 0.03$ \\
Kepler-470   &     11974540 & $6613 \pm  200$ & $4.26 \pm 0.07$ & $+0.04 \pm 0.14$ \\
Kepler-471   &      7778437 & $6733 \pm  288$ & $4.29 \pm 0.09$ & $+0.07 \pm 0.15$ \\
Kepler-485   &     12019440 & $5801 \pm   60$ & $4.42 \pm 0.03$ & $+0.19 \pm 0.04$ \\
Kepler-489   &      2987027 & $4832 \pm  123$ & $4.53 \pm 0.07$ & $-0.12 \pm 0.09$ \\
Kepler-490   &     10019708 & $6045 \pm  134$ & $4.25 \pm 0.06$ & $-0.02 \pm 0.15$ \\
Kepler-491   &      6849046 & $5521 \pm   60$ & $4.45 \pm 0.03$ & $+0.37 \pm 0.04$ \\
Kepler-492   &      7046804 & $5237 \pm   60$ & $4.56 \pm 0.03$ & $+0.14 \pm 0.12$ \\
Kepler-670   &     11414511 & $5709 \pm  111$ & $4.58 \pm 0.05$ & $+0.07 \pm 0.14$ \\
\hline
\noalign{\smallskip}
Star & 
  \multicolumn{1}{l}{TIC} &
  \multicolumn{1}{c}{T$_{\rm eff}$ [K] } &
  \multicolumn{1}{c}{$\log g$ } & 
  \multicolumn{1}{c}{[Fe/H]} \\
\hline
HD 271181    &    179317684 & $6495 \pm   90$ & $4.20 \pm 0.03$ & $+0.22 \pm 0.04$ \\
KELT-23      &    458478250 & $5899 \pm   49$ & $4.46 \pm 0.02$ & $-0.11 \pm 0.08$ \\
KELT-24      &    349827430 & $6509 \pm   50$ & $4.25 \pm 0.02$ & $+0.19 \pm 0.08$ \\
TOI-1181     &    229510866 & $6121 \pm   60$ & $4.23 \pm 0.10$ & $-0.08 \pm 0.06$ \\
TOI-1268     &    142394656 & $5290 \pm  117$ & $4.52 \pm 0.04$ & $+0.34 \pm 0.11$ \\
TOI-1296     &    219854185 & $5603 \pm   47$ & $4.10 \pm 0.02$ & $+0.44 \pm 0.04$ \\
WASP-18      &    100100827 & $6599 \pm   48$ & $4.43 \pm 0.02$ & $+0.22 \pm 0.03$ \\
WASP-62      &    149603524 & $6355 \pm   26$ & $4.39 \pm 0.01$ & $+0.23 \pm 0.02$ \\
WASP-100     &     38846515 & $6877 \pm   95$ & $4.19 \pm 0.03$ & $+0.01 \pm 0.06$ \\
WASP-126     &     25155310 & $5746 \pm   20$ & $4.33 \pm 0.01$ & $+0.14 \pm 0.02$ \\
\hline
	\end{tabular}
\end{table}

\subsection{Target selection}
I have selected stars observed by the {\it Kepler} \citep{2010Sci...327..977B} and {\it TESS} \citep{2015JATIS...1a4003R} missions for my analysis. I used the search tool\footnote{\url{ https://archive.stsci.edu/kepler/koi/search.php}} provided by the Mikulski Archive for Space Telescopes (MAST) to select Kepler objects of interest (KOIs) that are confirmed planets where the transit signal has a signal-to-noise ratio $>500$, orbital period $P<30$\,d, transit impact parameter $b<0.8$, and a host star with an effective temperature T$_{\rm eff} < 7000$\,K. 
 Hot stars were avoided because they shows complications in the light curve due to pulsations, gravity darkening, etc. 
 Planets with a high transit impact parameter were avoided because their light curves contain little information on the limb darkening of the host star \citep{2013A&A...560A.112M}. Short-period planets are preferred so that the light curve contains many transits. This avoid complications due to systematic errors in a few transits giving spurious results. 
 Stars known to show transits from multiple planets or transit timing variations were excluded from the sample. 
 I also excluded HAT-P-11 and Kepler-71 from the sample because their light curves are badly affected by the planet crossing star spots during the transit \citep{2011ApJ...743...61S,2019MNRAS.484..618Z}. 
 I used only short-cadence data for this analysis so stars with little or no short-cadence data were also excluded. The stars selected for analysis are listed in Table~\ref{tab:stars}. 
 
 I used the TEPCat catalogue of transiting extrasolar planets \citep{TEPCAT} and {\tt lightkurve}\footnote{\url{https://docs.lightkurve.org/}} to select stars brighter than $V=11.5$ showing transits at least 0.5\,per\,cent deep due to planets having an orbital period $P<10$\,d for which {\rm TESS} 2-minute light curves in at least 5 sectors are available from MAST. The stars selected for analysis are also listed in Table~\ref{tab:stars}.
 
 The values of T$_{\rm eff}$, $\log g $ and [Fe/H] for all stars are taken from the SWEET-Cat catalogue \citep{2021A&A...656A..53S}. Where possible, I used the $\log g$ value based on the data from the Gaia eDR3 catalogue \citep{2021A&A...649A...1G} since this is thought to be more reliable than the $\log g$ values based on spectroscopy (S. Sousa, priv.comm.). The $\log g$ value based on spectroscopy was used in a few cases where no $\log g$ value based on Gaia eDR3 data was available.

\begin{figure}
	\includegraphics[width=\columnwidth]{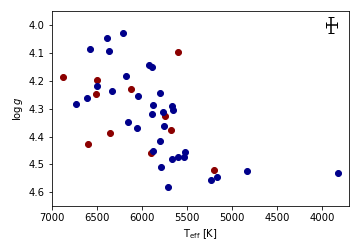}
    \caption{Selected target stars in the T$_{\rm eff}$ -- $\log g$ plane. Targets with Kepler or TESS light curves are plotted using blue and red symbols, respectively. The typical errors on the observed values is indicated by an error bar in the upper-right corner of the plot.} 
    \label{fig:teff_logg}
\end{figure}

 Fig.~\ref{fig:teff_logg} shows the selected target stars are shown in the T$_{\rm eff}$ -- $\log g$ plane. The sample is dominated by F- and G-type dwarf stars but there are also one or two K-type dwarfs in the Kepler sample.

% lcfit figure

\subsection{Pre-processing of the light curve data}\label{sec:preproc}
 I used the short-cadence pre-search data conditioning SAP fluxes (PDCSAP\_FLUX) provided in the {\it Kepler} archive files for my analysis. 
 Only data within one transit duration of the times of mid-transit were used for this analysis. The flux values for each transit were divided by a straight line fit to the flux values either side of the transit. The normalised fluxes from each quarter were then combined into a single phase-binned light curve in two steps. I first calculated the median value in phase bins of width 180\,s. I then rejected points more than 5 times the mean absolute deviation in each phase bin from the analysis. The remaining points were then used to calculate the mean and standard error of the mean in  phase bins of width 60\,s. The time value assigned to each phase bin corresponds to a time near the middle of the quarter at the same orbital phase  as the mean phase of the points in the bin. This allows me to include the orbital period as a free parameter in the fit to the data from all quarters.
 
 For the {\it TESS} data I used the same steps as for the {\it Kepler} data to normalize the fluxes and identify outliers. There is, in general, less data available for these targets so I do not phase-bin the data prior to further analysis. The error assigned to each data point is taken to be 1.25 times the mean absolute deviation of points in the same phase bin so that regions of the light curve that show excess noise are appropriately down-weighted in the analysis. A phase bin width of 120\,s was used for all these calculations. 
 
\subsection{Transit model fits}
 To model the transits in the light curves of the selected stars I used {\tt batman} version 2.4.7 \citep{2015PASP..127.1161K} with Claret's 4-parameter non-linear limb-darkening law. For the {\it TESS} data I used 3-point numerical integration to account for the exposure time of 120\,s.
 The free parameters in the fit were: orbital period, $P$; time of mid-transit, $T_0$; planet-star radius ratio, $k=R_{\rm pl}/R_{\star}$; host star radius relative to the orbital semi-major axis, $R_{\star}/a$; transit impact parameter $b = a\cos{i}/R_{\star}$ (where $i$ is the planet's orbital inclination); and the limb-darkening coefficients, $a_1,\dots,a_4$. For systems where I found an independent measurement of the orbital eccentricity that is significantly different from $e=0$, I also include $e$ and $\omega$ (the longitude of periastron) as free parameters but with Gaussian priors applied so that they remain consistent with the independently-measured values.
 For every trial set  of limb-darkening coefficients, $a_1,\dots,a_4$, the limb-darkening profile was computed at 100 uniformly-distributed values of $\mu$ from 0.01 to 1. This calculation was used to reject solutions where these coefficients do not correspond to a physically realistic limb-darkening profile, i.e. $0 < I(\mu) <1 $ and $\frac{dI(\mu)}{d\mu} > 0$ for all values of $\mu$. To sample the posterior probability distribution (PPD) for the vector of model parameters $\bmath{\theta}$ given the observed light curve, $D$, $P(\bmath{\theta} | D) \propto P(D|\bmath{\theta})P(\bmath{\theta})$ I used the affine-invariant Markov chain Monte Carlo sampler {\tt emcee} \citep{2013PASP..125..306F, GoodmanWeare2010}. To compute the likelihood $P(D|\bmath{\theta})$ I assume that the error on data point $i$ has a Gaussian distribution with standard deviation $f\sigma_i$ and that these errors are independent. The logarithm of the error-scaling factor $f$ is included as a hyper-parameter in the vector of model parameters $\bmath{\theta}$. I used uniform priors on all model parameters within the full range allowed by the {\tt batman} model.
 
 I used 100 walkers and 1000 steps to generate a random sample of points from the PPD following 4000 ``burn-in'' steps. Convergence of the chain was confirmed by visual inspection of the sample values for each parameter as a function of step number to ensure that there are no trends in the mean values or variances for the sample values from all walkers after the burn-in phase. 

\begin{figure*}
	\includegraphics[width=\textwidth]{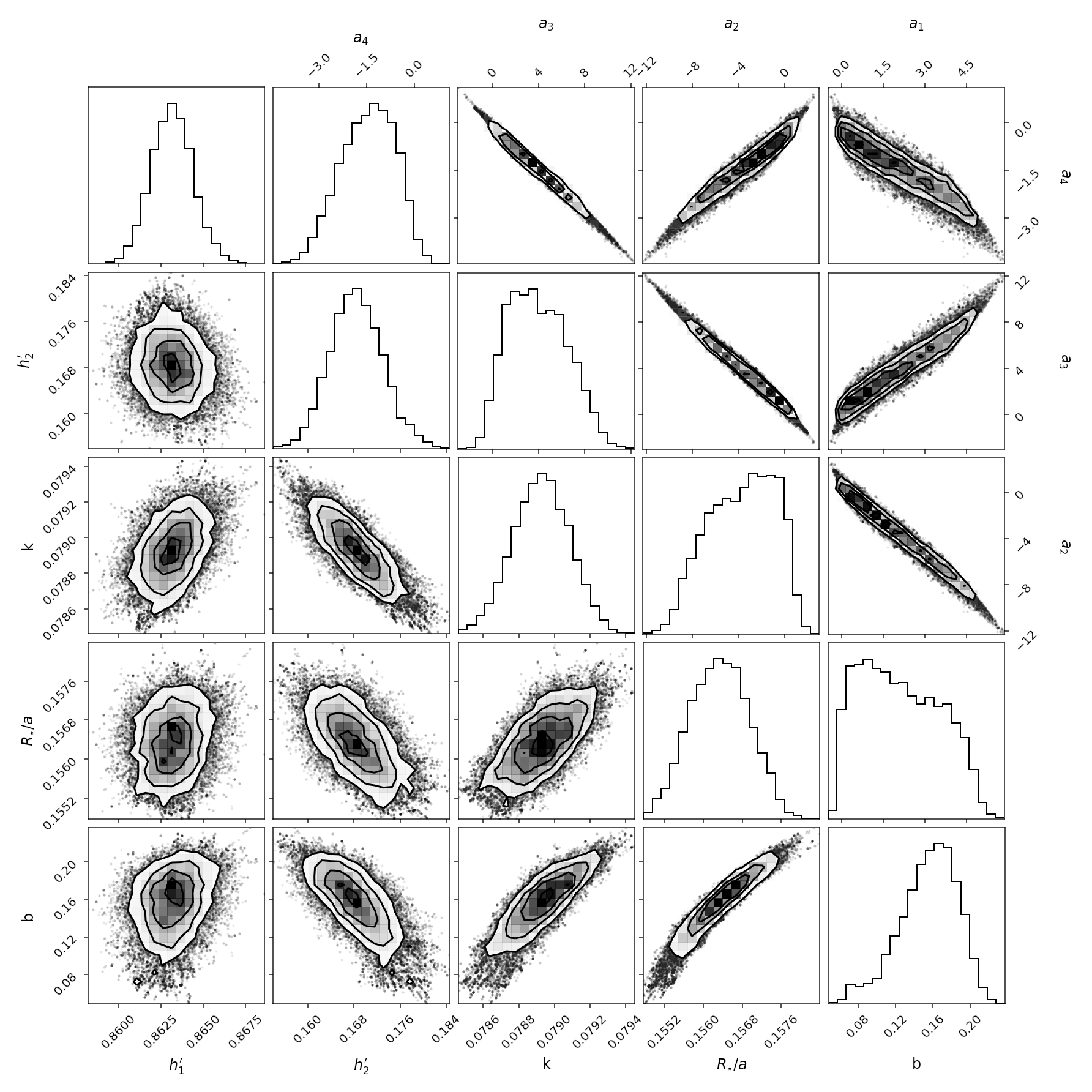}
    \caption{Parameter correlation plots from our analysis of the Kepler light curve of Kepler-5. This plot was produced using the package {\tt corner} \citep{corner}.}
    \label{fig:Kepler-15_corner}
\end{figure*}

A typical parameter correlation plot for selected parameters is shown in Fig.~\ref{fig:Kepler-15_corner}. The mean and standard deviation for the parameters of interest calculated from the sampled PPD are summarised in Table~\ref{tab:results}. Note that the orbital inclination is allowed to exceed $i=90^{\circ}$ so the PPD may include negative values of $b$. 
Examples of the best fits to typical {\it Kepler} and {\it TESS} light curves are shown in Fig.~\ref{fig:lcfit}. Also shown in Fig.~\ref{fig:lcfit} is the fit to the light curve of Kepler-17, a star that shows a moderate level of magnetic activity, resulting in excess scatter through the transit. To measure this excess scatter I computed the ratio of the standard deviation of the residuals in the bottom half of the transit to the standard deviation of the data outside the transit. This ratio, $r$, is also given in Table~\ref{tab:results}. Note that the planet parameters in Table~\ref{tab:results} do not account for tidal deformation of the planet \citep{2014ApJ...789..113B, 2014A&A...570L...5C}.

I also analysed all the light curves using the polynomial limb-darkening law given in equation~{\ref{eqn:poly}}. The coefficients $c_5$ and $c_6$ were both fixed to $0$ but the other 4 coefficients were included as free parameters in the fit. The details of the analysis are otherwise identical to those described above. The results are almost identical to those obtained using Claret's 4-parameter limb-darkening so are not reported here. This does demonstrate that the conclusions of this study are not affected by the choice of limb-darkening law used to analyse the light curves.
 
\subsection{Notes on individual objects}

\subsubsection{Kepler-6}
A correction for a small amount of third light, $\ell_3 = 0.033$, reported by \citet{2010ApJ...713L.136D} was made prior to analysis.

\subsubsection{Kepler-14}
I used priors on $e=0.035 \pm  0.017$ and $\omega= 89^{\circ} \pm 16^{\circ}$ derived from the spectroscopic orbit by \citet{2011ApJS..197....3B}.

\subsubsection{Kepler-423}
 I used priors on $e=0.019\,^{+0.028}_{-0.014}$ and $\omega= 120^{\circ}\,^{+77^{\circ}}_{-34^{\circ}}$ derived from the  spectroscopic orbit by \citet{2015A&A...576A..11G}. 
 
\subsubsection{Kepler-489}
 {\it Kepler} short-cadence observations of Kepler-489 in Quarter 7 only cover two transits of Kepler-489\,b so the data from this quarter were excluded from our analysis.
 
 \subsubsection{HD 271181 (TOI-163)}
 The ephemeris for the time of mid-transit published in \citet{2019MNRAS.490.1094K} is inconsistent with the following ephemeris that I obtained from the analysis of the {\it TESS} photometry (figures in parentheses are standard errors in the final digit):
 \[{\rm BJD}_{\rm TDB} = 2459149.7199 (1) +  4.231115 (1)\cdot E.\]
 It is also inconsistent with the ephemeris listed in the {\it TESS} ``objects of interest'' catalogue published on 2021-12-09\footnote{\url{https://exoplanetarchive.ipac.caltech.edu}} that I used for pre-processing the data. If I use the ephemeris from \citeauthor{2019MNRAS.490.1094K} to phase-fold the TESS data, I find that the transits occur approximately 40 minutes too early.

% Kepler-5_corner figure

% lcfit figure
\begin{figure}
	\includegraphics[width=\columnwidth]{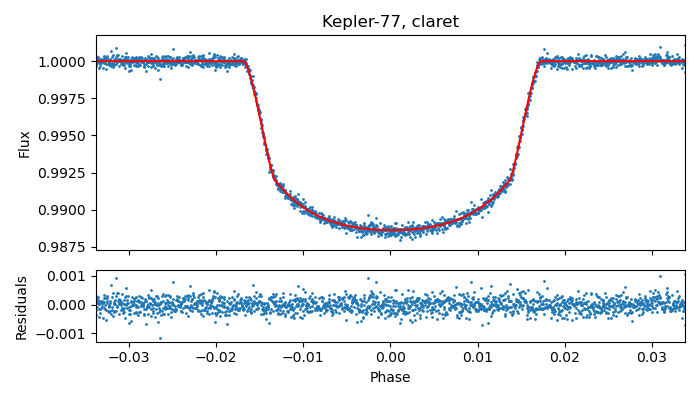}
	\includegraphics[width=\columnwidth]{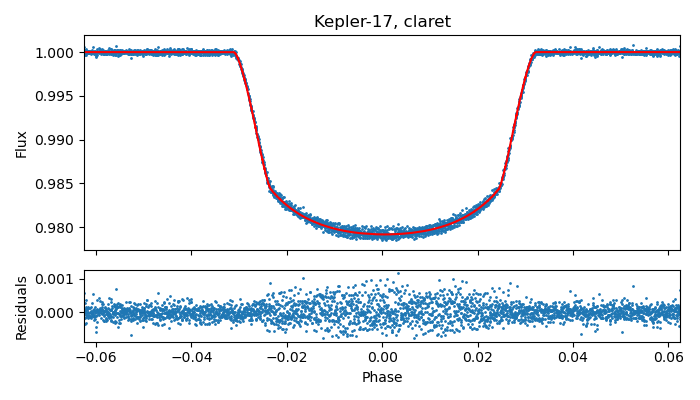}
	\includegraphics[width=\columnwidth]{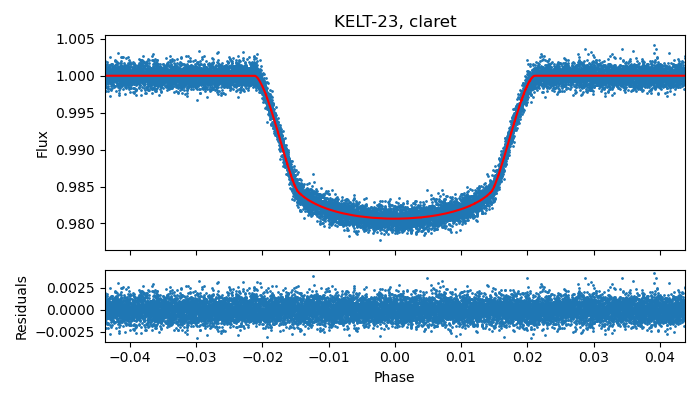}
    \caption{Light curves for three stars together with the best fit from our analysis (red line). The light curve for KELT-23 is from TESS, the other two light curves are from Kepler and have been phase-binned prior to analysis.} 
    \label{fig:lcfit}
\end{figure}

% Results table
\begin{table*}
\centering
\caption{Results from our light curve analysis. Figures in parentheses give the standard error in the preceding digit. $C(h^{\prime}_1,h^{\prime}_2)$ is the correlation coefficient for the parameters $h^{\prime}_1$ and $h^{\prime}_2$. The number of points in the light curve and the standard deviation of the residuals are given in the columns headed $N$ and $\sigma$, respectively. Results in the first part of the table are based on {\it Kepler} data and results in the second part of the table are based on {\it TESS} data.}
	\label{tab:results}
 \addtolength\tabcolsep{-0.25pt} % Slight squeeze on column widths

\begin{tabular}{@{}lrrrrrrrrrrr} 
\hline		
Star & 
  \multicolumn{1}{c}{P [d]} &
  \multicolumn{1}{c}{$h^{\prime}_1$ } & 
  \multicolumn{1}{c}{$h^{\prime}_2$ } &
  \multicolumn{1}{c}{$C(h^{\prime}_1,h^{\prime}_2)$ } & 
  \multicolumn{1}{c}{$k=R_{\rm pl}/R_{\star}$} &
  \multicolumn{1}{c}{$R_{\star}/a$} &
  \multicolumn{1}{c}{$b$} &
  \multicolumn{1}{c}{$N$} &
  \multicolumn{1}{c}{$\sigma$} [ppm]& $r$& $f$ \\ 
		\hline
HAT-P-7      &  2.20   &$    0.859 \pm    0.001   $&$    0.177 \pm    0.001 $&$ +0.09   $&$  0.07738 (6)   $&$   0.2407 (2)   $&$    0.491 (2) $ &  8357  &   41  & 1.35 & 1.30  \\
Kepler-4     &  3.21   &$     0.86 \pm     0.01   $&$     0.23 \pm     0.01 $&$ +0.49   $&$   0.0247 (2)   $&$    0.168 (6)   $&$      0.3 (1) $ &  4611  &   93  & 0.99 & 1.09  \\
Kepler-5     &  3.55   &$    0.863 \pm    0.001   $&$    0.168 \pm    0.005 $&$ -0.15   $&$   0.0789 (2)   $&$   0.1563 (6)   $&$     0.15 (3) $ &  6078  &  144  & 1.00 & 1.06  \\
Kepler-6     &  3.23   &$    0.822 \pm    0.001   $&$    0.214 \pm    0.006 $&$ -0.33   $&$   0.0915 (2)   $&$   0.1339 (5)   $&$     0.15 (3) $ &  3508  &  120  & 1.16 & 1.04  \\
Kepler-7     &  4.89   &$    0.850 \pm    0.008   $&$    0.192 \pm    0.004 $&$ -0.04   $&$   0.0822 (2)   $&$   0.1504 (5)   $&$    0.554 (6) $ &  3748  &  121  & 1.05 & 1.08  \\
Kepler-8     &  3.52   &$     0.85 \pm     0.03   $&$    0.178 \pm    0.009 $&$ +0.81   $&$   0.0946 (4)   $&$   0.1466 (4)   $&$    0.719 (4) $ &  4281  &  198  & 1.05 & 1.05  \\
Kepler-12    &  4.44   &$   0.8438 \pm   0.0009   $&$    0.187 \pm    0.005 $&$ -0.29   $&$   0.1176 (2)   $&$   0.1251 (3)   $&$     0.18 (1) $ &  5067  &  203  & 1.04 & 1.15  \\
Kepler-14    &  6.79   &$     0.86 \pm     0.01   $&$    0.173 \pm    0.004 $&$ +0.71   $&$   0.0454 (1)   $&$    0.140 (3)   $&$     0.62 (2) $ &  7955  &   95  & 1.02 & 1.11  \\
Kepler-15    &  4.94   &$     0.85 \pm     0.02   $&$    0.220 \pm    0.010 $&$ +0.64   $&$   0.1025 (4)   $&$   0.1017 (4)   $&$    0.683 (5) $ &  1988  &  232  & 1.00 & 1.11  \\
Kepler-17    &  1.49   &$    0.841 \pm    0.001   $&$    0.165 \pm    0.007 $&$ -0.30   $&$   0.1328 (3)   $&$   0.1769 (4)   $&$     0.18 (2) $ &  2951  &  255  & 1.96 & 1.36  \\
Kepler-41    &  1.86   &$     0.87 \pm     0.02   $&$    0.214 \pm    0.010 $&$ +0.57   $&$   0.1003 (4)   $&$   0.1964 (9)   $&$    0.685 (6) $ &  1151  &  210  & 0.94 & 1.05  \\
Kepler-43    &  3.02   &$     0.87 \pm     0.02   $&$    0.206 \pm    0.006 $&$ +0.50   $&$   0.0854 (2)   $&$   0.1446 (6)   $&$    0.658 (5) $ &  3795  &  174  & 1.04 & 1.04  \\
Kepler-44    &  3.25   &$     0.83 \pm     0.04   $&$     0.20 \pm     0.02 $&$ +0.58   $&$   0.0807 (7)   $&$    0.144 (3)   $&$     0.64 (2) $ &  1463  &  454  & 1.00 & 1.09  \\
Kepler-45    &  2.46   &$     0.81 \pm     0.02   $&$     0.23 \pm     0.02 $&$ +0.40   $&$   0.1814 (9)   $&$   0.0934 (4)   $&$    0.560 (6) $ &  2420  &  721  & 1.04 & 1.08  \\
Kepler-74    &  7.34   &$     0.82 \pm     0.05   $&$     0.18 \pm     0.02 $&$ +0.73   $&$    0.091 (1)   $&$   0.0657 (8)   $&$     0.70 (1) $ &  1440  &  442  & 1.12 & 1.26  \\
Kepler-77    &  3.58   &$    0.827 \pm    0.004   $&$    0.201 \pm    0.007 $&$ -0.30   $&$   0.0984 (3)   $&$   0.1038 (6)   $&$     0.38 (2) $ &  1746  &  228  & 0.99 & 1.09  \\
Kepler-412   &  1.72   &$     0.77 \pm     0.08   $&$     0.19 \pm     0.02 $&$ +0.96   $&$    0.104 (2)   $&$    0.205 (1)   $&$    0.792 (4) $ &   992  &  194  & 0.92 & 1.04  \\
Kepler-422   &  7.89   &$    0.838 \pm    0.006   $&$    0.198 \pm    0.006 $&$ -0.27   $&$   0.0957 (2)   $&$   0.0737 (3)   $&$    0.492 (7) $ &  6250  &  233  & 0.98 & 1.11  \\
Kepler-423   &  2.68   &$    0.829 \pm    0.002   $&$    0.209 \pm    0.008 $&$ -0.43   $&$   0.1237 (3)   $&$    0.125 (3)   $&$     0.33 (1) $ &  1944  &  210  & 0.97 & 1.08  \\
Kepler-425   &  3.80   &$     0.81 \pm     0.02   $&$     0.23 \pm     0.02 $&$ +0.21   $&$   0.1144 (7)   $&$   0.0863 (6)   $&$     0.60 (1) $ &  1120  &  329  & 0.89 & 1.08  \\
Kepler-426   &  3.22   &$     0.85 \pm     0.04   $&$     0.22 \pm     0.02 $&$ +0.75   $&$    0.118 (1)   $&$   0.1046 (6)   $&$    0.722 (6) $ &  1269  &  391  & 1.00 & 1.09  \\
Kepler-427   & 10.29   &$    0.826 \pm    0.005   $&$     0.20 \pm     0.02 $&$ +0.00   $&$   0.0896 (6)   $&$   0.0508 (6)   $&$     0.15 (9) $ &  2040  &  436  & 1.07 & 1.48  \\
Kepler-433   &  5.33   &$    0.860 \pm    0.006   $&$     0.17 \pm     0.01 $&$ +0.17   $&$   0.0633 (3)   $&$    0.144 (2)   $&$     0.15 (8) $ &  2975  &  393  & 0.99 & 1.15  \\
Kepler-435   &  8.60   &$     0.87 \pm     0.01   $&$     0.19 \pm     0.01 $&$ +0.02   $&$   0.0627 (3)   $&$    0.140 (2)   $&$     0.42 (3) $ &  3228  &  259  & 0.97 & 1.15  \\
Kepler-470   & 24.67   &$    0.875 \pm    0.009   $&$     0.15 \pm     0.01 $&$ +0.34   $&$   0.0806 (2)   $&$   0.0375 (5)   $&$     0.44 (2) $ &  1724  &  367  & 0.99 & 1.66  \\
Kepler-471   &  5.01   &$     0.87 \pm     0.01   $&$     0.17 \pm     0.01 $&$ -0.08   $&$   0.0765 (4)   $&$    0.122 (2)   $&$     0.41 (3) $ &  1648  &  310  & 1.10 & 1.21  \\
Kepler-485   &  3.24   &$    0.834 \pm    0.003   $&$     0.17 \pm     0.01 $&$ -0.28   $&$   0.1184 (5)   $&$   0.1116 (7)   $&$     0.22 (4) $ &  1829  &  419  & 1.00 & 1.07  \\
Kepler-489   & 17.28   &$    0.785 \pm    0.006   $&$     0.20 \pm     0.02 $&$ +0.13   $&$    0.092 (1)   $&$   0.0280 (5)   $&$      0.2 (1) $ &  1423  &  468  & 0.98 & 1.29  \\
Kepler-490   &  3.27   &$    0.850 \pm    0.005   $&$     0.19 \pm     0.02 $&$ -0.34   $&$   0.0927 (5)   $&$    0.131 (1)   $&$     0.27 (5) $ &  1667  &  337  & 0.98 & 1.05  \\
Kepler-491   &  4.23   &$    0.818 \pm    0.010   $&$     0.23 \pm     0.01 $&$ -0.19   $&$   0.0803 (5)   $&$    0.090 (1)   $&$     0.44 (3) $ &  1363  &  230  & 1.04 & 1.08  \\
Kepler-492   & 11.72   &$     0.80 \pm     0.06   $&$     0.19 \pm     0.03 $&$ +0.84   $&$    0.097 (1)   $&$   0.0403 (4)   $&$     0.70 (1) $ &  1446  &  538  & 1.04 & 1.62  \\
Kepler-670   &  2.82   &$    0.821 \pm    0.007   $&$     0.21 \pm     0.02 $&$ -0.48   $&$   0.1197 (7)   $&$   0.1136 (7)   $&$     0.41 (2) $ &  1805  &  416  & 0.98 & 1.04  \\
TrES-2       &  2.47   &$     0.89 \pm     0.03   $&$    0.201 \pm    0.006 $&$ +0.82   $&$   0.1242 (6)   $&$   0.1275 (4)   $&$    0.845 (1) $ &  3179  &   60  & 1.13 & 1.18  \\
\noalign{\smallskip}\hline\noalign{\smallskip}
HD 271181    &  4.23   &$    0.901 \pm    0.009   $&$     0.17 \pm     0.02 $&$ +0.04   $&$   0.0805 (6)   $&$    0.130 (3)   $&$     0.30 (7) $ &  24231  & 1656  & 1.01 & 1.01 \\
KELT-23      &  2.26   &$     0.86 \pm     0.01   $&$     0.17 \pm     0.01 $&$ -0.24   $&$   0.1326 (4)   $&$   0.1319 (5)   $&$    0.524 (8) $ &  13912  &  897  & 0.99 & 1.00  \\
KELT-24      &  5.55   &$    0.893 \pm    0.002   $&$    0.142 \pm    0.007 $&$ +0.14   $&$   0.0870 (2)   $&$   0.0936 (4)   $&$     0.08 (5) $ &   8781  &  404  & 1.06 & 1.03  \\
TOI-1181     &  2.10   &$    0.875 \pm    0.006   $&$     0.18 \pm     0.02 $&$ +0.10   $&$   0.0762 (5)   $&$    0.245 (4)   $&$     0.27 (7) $ &  27360  & 1080  & 1.03 & 1.01  \\
TOI-1268     &  8.16   &$    0.860 \pm    0.009   $&$     0.20 \pm     0.02 $&$ -0.01   $&$   0.0898 (6)   $&$    0.058 (1)   $&$      0.2 (1) $ &   4033  & 1127  & 1.11 & 1.03  \\
TOI-1296     &  3.94   &$     0.85 \pm     0.01   $&$     0.18 \pm     0.02 $&$ +0.15   $&$   0.0760 (7)   $&$    0.154 (4)   $&$      0.2 (1) $ &  16549  & 1575  & 0.99 & 1.02  \\
WASP-18      &  0.94   &$    0.875 \pm    0.005   $&$     0.15 \pm     0.01 $&$ -0.28   $&$   0.0970 (3)   $&$    0.288 (1)   $&$     0.37 (2) $ &  14349  &  588  & 1.09 & 1.00  \\
WASP-62      &  4.41   &$    0.889 \pm    0.003   $&$    0.133 \pm    0.009 $&$ -0.31   $&$   0.1110 (3)   $&$   0.1034 (4)   $&$     0.25 (2) $ &  31427  &  921  & 1.01 & 1.01  \\
WASP-100     &  2.85   &$     0.88 \pm     0.02   $&$     0.14 \pm     0.01 $&$ +0.13   $&$   0.0824 (3)   $&$    0.187 (1)   $&$     0.56 (1) $ &  50273  & 1220  & 1.03 & 1.01  \\
WASP-126     &  3.29   &$    0.864 \pm    0.005   $&$     0.20 \pm     0.02 $&$ -0.21   $&$   0.0771 (4)   $&$    0.128 (1)   $&$     0.11 (8) $ &  39099  & 1427  & 1.01 & 1.01  \\
\noalign{\smallskip}\hline
	\end{tabular}
\end{table*}

\section{Discussion}\label{sec:discussion}

\subsection{Estimating the mean offset accounting for additional scatter}\label{sec:combine}
 In the following discussion we frequently wish to measure an offset between estimates of $h^{\prime}_1$ and $h^{\prime}_2$ from two different sources. I assume that the measurements of $h^{\prime}_1$ and $h^{\prime}_2$ have some extra scatter beyond their quoted standard errors. This extra scatter, $\sigma_{\rm ext}$, will be a combination of variance of astrophysical origin, e.g. due to magnetic activity on the star, and systematic errors e.g. imperfect removal of instrumental noise. If we assume that all errors are independent and have a Gaussian distribution then the log-likelihood to obtain the observed difference $\bmath{\Delta } = \{\Delta_i \pm \sigma_i, i=1,\dots,N\}$ is 
 \[\ln\,p(\bmath{\Delta}\,|\,\langle\Delta\rangle,\sigma_{\rm ext}) = -\frac{1}{2} \sum_i \left[
    \frac{(\Delta_i-\langle\Delta\rangle)^2}{s_i^2}
    + \ln \left ( 2\pi\,s_i^2 \right )
\right], \]
where $s_i^2 = \sigma_i^2 + \sigma_{\rm ext}^2$. I assume a broad uniform prior on the mean offset, $\langle\Delta\rangle$, and a broad uniform  prior on $\ln \sigma_{\rm ext}$. I then sample the posterior probability distribution using {\tt emcee} with 1500 steps and 128 walkers. After discarding the first 500 ``burn-in'' steps of the Markov chain, I use the remaining sample to calculate the mean and standard deviation of the posterior probability distribution for $\langle\Delta\rangle$, i.e. the best estimate for the value of the offset and its standard error.

% compare_h1h2_maxted_claret figure
\begin{figure}
	\includegraphics[width=\columnwidth]{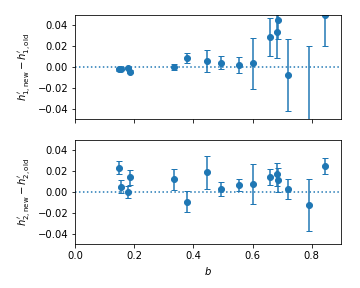}
    \caption{Difference between the values of $h^{\prime}_1$ and $h^{\prime}_2$ measured from {\it Kepler} light curves by \citet{2018A+A...616A..39M} and from this study. }
    \label{fig:compare_h1ph2p_maxted_claret}
\end{figure}

\subsection{Comparison to Maxted (2018)}
 There are 16 stars in common between this study and \citet{2018A+A...616A..39M}, in which I analysed {\it Kepler} light curves assuming a power-2 limb-darkening law. These studies also differ in the way that the light curve data were processed prior to analysis, e.g. phase-binning, outlier rejection and normalisation, and the details of the analysis such as the assignment of weights to the data, the model used to analyse the data ({\tt ellc} versus {\tt batman}), etc. Fig.~\ref{fig:compare_h1ph2p_maxted_claret} shows the difference between the values of $h^{\prime}_1$ and $h^{\prime}_2$ between these two studies. The agreement between the values of $h^{\prime}_1$ from these two studies is excellent ($\Delta h^{\prime}_1 = 0.000 \pm 0.008$, $\sigma_{\rm ext,1} = 0.001$). There is a small offset in the values of $h^{\prime}_2$ between these two studies ($\Delta h^{\prime}_2 = 0.010 \pm 0.002$, $\sigma_{\rm ext,2} = 0.001$). I repeated the analysis of the phase-binned {\it Kepler} light curves described in Section~\ref{sec:preproc} for these 16 stars using a power-2 law instead of the Claret 4-parameter law. The results are very similar for the differences between the values of $h^{\prime}_1$ and $h^{\prime}_2$. The implication of these results is that the measured values of  $h^{\prime}_1$ are robust but the values of $h^{\prime}_2$ may be affected by systematic errors $\approx 0.01$ depending on the details of the analysis.

% compare_h1ph2p_kostogryz_claret_set2 figure
\begin{figure}
	\includegraphics[width=\columnwidth]{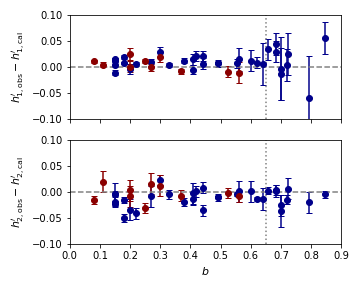}
   \caption{Comparison of observed values of $h^{\prime}_1$ and $h^{\prime}_2$ to predicted values calculated using the  ``Set 2'' models from   \citet{2022arXiv220606641K} as a function of impact parameter, $b$. Points for stars observed using {\it Kepler} and {\it TESS} are colour-coded blue and red, respectively.}
    \label{fig:compare_h1ph2p_kostogryz_claret_set2}
\end{figure}

% compare_h1ph2p_phoenix_claret_teff figure
\begin{figure}
	\includegraphics[width=\columnwidth]{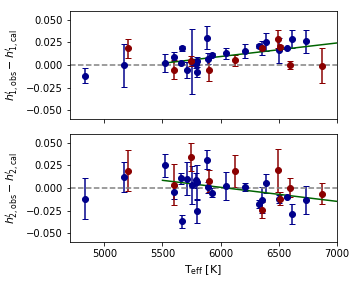}
   \caption{Comparison of observed values of $h^{\prime}_1$ and $h^{\prime}_2$ to predicted values calculated using PHOENIX-COND models \citep{2018A+A...618A..20C}  for systems with impact parameter $b<0.65$. Kepler-45 with $T_{\rm eff} = 3820$\,K is not shown here. Points for stars observed using {\it Kepler} and {\it TESS} are colour-coded blue and red, respectively. The linear fits to the residuals for stars with $T_{\rm eff} > 5500$\,K} described in the text are shown in green.
    \label{fig:compare_h1ph2p_phoenix_claret_teff}
\end{figure}

\subsection{Comparison to limb-darkening profiles from models}

\subsubsection{Technical details}
For each set of tabulated limb-darkening coefficients, the predicted values of $h^{\prime}_1$ and $h^{\prime}_2$ for each star are computed by linear interpolation based on the values of $T_{\rm eff}$, $\log g$ and [Fe/H] given in Table~\ref{tab:stars}. Errors on $h^{\prime}_1$ and $h^{\prime}_2$ are computed using a Monte Carlo method assuming Gaussian independent errors on these input values. For most models, the interpolation yields values of the coefficients $a_1,\dots, a_4$ that are then used to compute $h^{\prime}_1$ and $h^{\prime}_2$ using equation~(\ref{eqn:claret}). For the comparison to the results of \citet{2022arXiv220606641K} I use the tabulated values of $I_{\lambda}(\mu)$ directly with linear interpolation to compute $h^{\prime}_1$ and $h^{\prime}_2$. \citet{2022arXiv220606641K} provide two sets of limb-darkening profiles, ``Set 1'' with a fixed value of the mixing-length parameter and chemical abundances relative to the solar composition from \citet{1998SSRv...85..161G}, and ``Set 2'' using a variable mixing-length parameter depending on T$_{\rm eff}$ and the solar composition from \citet{2009ARA&A..47..481A}. The difference between the observed and calculated values of $h^{\prime}_1$ and $h^{\prime}_2$ for ``Set 2'' as a function of impact parameter, $b$, are shown in Fig.~\ref{fig:compare_h1ph2p_kostogryz_claret_set2}.

The mean offset and external scatter for each set of coefficients computed in the sense ``observed $-$ calculated'' using the method described in Section~\ref{sec:combine} are given in Table~\ref{tab:results2}. Based on the simulations described in Section~\ref{sec:simulations}, I have only used systems with measured impact parameters $b < 0.65$ for this comparison. This is to avoid systems with strongly correlated values of $h^{\prime}_1$ and $h^{\prime}_2$ ($C(h^{\prime}_1,h^{\prime}_2) \goa 0.5$). 
For systems with $b < 0.65$, $h^{\prime}_1$ and $h^{\prime}_2$ will be correlated but there are similar numbers of stars with positive and negative values of $C(h^{\prime}_1,h^{\prime}_2)$ so the statistics in Table~\ref{tab:results2} should be reliable.

The limb-darkening profiles for PHOENIX-COND models from \citet{2018A+A...618A..20C} are calculated on a grid that extends beyond the limb so the limb-darkening coefficients in these tables cannot be used directly. Claret defines the limb to occur at $\mu^{\prime} = \mu_{\rm cri}$ and set $I_{\lambda}(\mu^{\prime}) = 0 $ for $\mu^{\prime} > \mu_{\rm cri}$. To calculate $h^{\prime}_1$ and $h^{\prime}_2$ the independent variable must be re-scaled using
\[ \mu = (\mu^{\prime}-\mu_{\rm cri})/(1-\mu_{\rm cri}), \]
so $\mu = \nicefrac{2}{3}$ corresponds to $\mu^{\prime} = \nicefrac{2}{3}+\nicefrac{1}{3}\mu_{\rm cri}$ and  $\mu = \nicefrac{1}{3}$ corresponds to  $\mu^{\prime} = \nicefrac{1}{3} +  \nicefrac{2}{3}\mu_{\rm cri}$. 
Note that the values of $h^{\prime}_1$ and $h^{\prime}_2$ derived are insensitive to the details of how $\mu_{\rm cri}$ is calculated. Referring to Fig.~\ref{fig:CLVPlot} we see that the sharp drop in flux near the limb occurs over a narrow range between $r\approx 0.9982$ and  $r\approx 0.9998$. Taking the radius of the limb to be $r=0.9985 \pm 0.0003$, this corresponds to $\mu_{\rm cri} = 0.0548 \pm 0.0055$. This uncertainty on the value of $\mu_{\rm cri}$ results in errors of only $\pm 0.0008$ for $h^{\prime}_1$ and $\pm 0.0002$ for $h^{\prime}_2$. 

The limb-darkening coefficients for PHOENIX-COND models from \citet{2018A+A...618A..20C} and from MARCS models by \citet{2022RNAAS...6..248M} are only available at solar-metallicity. I have used the limb-darkening coefficients from Set 1 of \citet{2022arXiv220606641K} to calculate a linear correction for [Fe/H] to the values of $h^{\prime}_1$ and $h^{\prime}_2$ from these models. The mean values of $T_{\rm eff}$, $\log g$ and [Fe/H] for the stars in our sample are 
$\langle T_{\rm eff}\rangle = 6355$\,K, 
$\langle \log g\rangle = 4.39$ and  $\langle{\rm [Fe/H]}\rangle = 0.23$. For these values of $T_{\rm eff}$ and $\log g$, assuming ${\rm [Fe/H]}=0$ results in a value of $h^{\prime}_1$ that is 0.0027 too high and a value of $h^{\prime}_2$ that is 0.0035 too low in the TESS band cf. the values obtained assuming ${\rm [Fe/H]}=0.23$. In the Kepler band, $h^{\prime}_1$ is 0.0034 too high and $h^{\prime}_2$ is 0.0041 too low. This correction for metallicity has a very small influence on the results presented in Table~\ref{tab:results2}.

\subsubsection{Results}

Apart from the models by \citet{2013A+A...556A..86N}, the results in Table~\ref{tab:results2} are fairly similar for all models. Typically, there is a small but significant offset $\Delta h^{\prime}_1 \approx 0.006$ for both the TESS and Kepler bands. The offset $\Delta h^{\prime}_2 \approx -0.01$ typically seen for these models is probably not significant because is comparable to the systematic error due to differences in the analysis methods used discussed in the previous section.

\citet{2013A+A...556A..86N} found a large difference in the limb-darkening profiles they computed using plane-parallel (ATLAS) and spherically-symmetric (sATLAS) model atmospheres. This difference can be seen in Fig.~\ref{fig:CLVPlot}. They conclude that ``sphericity is important even for dwarf model atmospheres, leading to significant differences in the predicted coefficients''. The PHOENIX-COND models from \citet{2018A+A...618A..20C} also assume spherical symmetry. In contrast to \citeauthor{2013A+A...556A..86N}, I see very good agreement between the results from the PHOENIX-COND models and plane-parallel models, and good agreement between these models and the observations. There is rather poor agreement between the observed values of $h^{\prime}_1$ and $h^{\prime}_2$ and the predicted values from  \citeauthor{2013A+A...556A..86N} for both the ATLAS and sATLAS models. The conclusion regarding spherically-symmetric versus plane-parallel models from \citet{2013A+A...556A..86N} cannot be regarded as reliable until the poor agreement between their computed limb-darkening coefficients and results presented here is better understood.  

\subsubsection{Trends with effective temperature}
For the limb-darkening coefficients based on PHOENIX-COND models \citep{2018A+A...618A..20C} there is a clear trend in $\Delta h^{\prime}_1 = h^{\prime}_{\rm 1,obs} - h^{\prime}_{\rm 1,cal}$ with $T_{\rm eff}$ for stars with $T_{\rm eff}>5500$.  The corresponding trend for $\Delta h^{\prime}_2$ with $T_{\rm eff}$ is marginally significant. Fitting these trends for stars with  $T_{\rm eff}>5500$ observed with {\it Kepler} and {\it TESS} together I find
\begin{eqnarray*}
\Delta h^{\prime}_1 = (0.0095 \pm 0.0019) + (+0.015 \pm 0.005)\,Y,\\
\Delta h^{\prime}_2 = (0.0008 \pm 0.0031) + (-0.015 \pm 0.008)\,Y,
\end{eqnarray*}

where $Y = (T_{\rm eff}-6000\,{\rm K})/1000\,{\rm K}$.
These trends are shown in Fig.~\ref{fig:compare_h1ph2p_phoenix_claret_teff}. To achieve a fit with $\chi^2 = N_{\rm df}$ for these least-squares fit I added 0.00966 and 0.01638 in quadrature to the standard error estimates on $\Delta h^{\prime}_1$ and $\Delta h^{\prime}_2$, respectively.
There are only a few stars with $T_{\rm eff}<5500$\,K in our sample, so it is not clear if these trends continue to cooler stars. Similar trends in $\Delta h^{\prime}_1$ and $\Delta h^{\prime}_2$ with  $T_{\rm eff}$ are seen for the limb-darkening coefficients published by \citet{2013A+A...556A..86N}. For the sATLAS stellar models these trends are
\begin{eqnarray*}
\Delta h^{\prime}_1 = (+0.0374 \pm 0.0017) + (+0.025 \pm 0.005)\,Y,\\ 
\Delta h^{\prime}_2 = (-0.0106 \pm 0.0036) + (-0.017 \pm 0.010)\,Y.
\end{eqnarray*}
For the plane-parallel ATLAS models these trends are
\begin{eqnarray*}
\Delta h^{\prime}_1 = (-0.0219 \pm 0.0017) + (+0.034 \pm 0.005)\,Y,\\
\Delta h^{\prime}_2 = (+0.0372 \pm 0.0035) + (-0.024 \pm 0.010)\,Y.
\end{eqnarray*}

The limb-darkening coefficients for the Kepler band recently published by \citet{2022RNAAS...6..248M} also show trends in  $\Delta h^{\prime}_1$ and $\Delta h^{\prime}_2$ with $T_{\rm eff}$. For stars with $T_{\rm eff}>5500$ I find the following linear fits to these trends:
\begin{eqnarray*}
\Delta h^{\prime}_1 = (+0.0293 \pm 0.0016) + (+0.018 \pm 0.005)\,Y,\\ 
\Delta h^{\prime}_2 = (-0.0241 \pm 0.0036) + (-0.020 \pm 0.009)\,Y.
\end{eqnarray*}
To achieve a fit with $\chi^2 = N_{\rm df}$ for these least-squares fit I added 0.0048 and 0.0133 in quadrature to the standard error estimates on $\Delta h^{\prime}_1$ and $\Delta h^{\prime}_2$, respectively.
 
 No other models show any significant trend in $\Delta h^{\prime}_1$ or $\Delta h^{\prime}_2$ with $T_{\rm eff}$.

\subsubsection{Impact of magnetic activity}
In \citet{2018A+A...616A..39M}, I suggested that the small offsets between the observed values of $h_1$ and $h_2$ and the values predicted by the Stagger-grid models may be due to weak magnetic fields in the atmospheres of the solar-type stars studied. None of the atmosphere models discussed here, including the Stagger-grid models, include the impact of magnetic fields. The comparison of observed $h_2$ values to models is not straightforward for the reasons discussed above, but the conclusion regarding $h_1$ from \citet{2018A+A...616A..39M} is robust and very similar to the results for $h^{\prime}_1$ in this study.

The impact of a magnetic field on the limb-darkening of the Sun can be seen in Fig.~4 of \citet{2017A&A...605A..45N}. This figure shows the limb-darkening of a Sun-like star computed using the MURaM stellar atmosphere model assuming either zero magnetic field or including a mean vertical magnetic field strength of 100\,G, which is typical for the quiet Sun. The effect of this magnetic field at a wavelength of 611\,nm is to increase $h^{\prime}_1$ by 0.007 and to decrease $h^{\prime}_2$ by 0.005. This agrees very well with the observed values of $h^{\prime}_1$ and $h^{\prime}_2$ observed in the {\it Kepler} bandpass with a mean wavelength $\approx 630$\,nm. The tendency for the magnetic field to have less of an effect at redder wavelengths seen in Fig.~4 of \citeauthor{2017A&A...605A..45N} is also reflected in our results for the {\it TESS} bandpass with a mean wavelength $\approx 800$\,nm cf. the results for the bluer {\it Kepler} bandpass. 

An intriguing piece of evidence in favour of this interpretation is the case of WASP-18. This star has a significantly lower value of $h^{\prime}_1$ compared to stars with similar $T_{\rm eff}$. This can be seen in  Fig.~\ref{fig:compare_h1ph2p_phoenix_claret_teff}, where WASP-18 is the outlier that sits below the trend at $T_{\rm eff}\approx 6500$\,K and $\Delta h^{\prime}_1 \approx 0$, i.e. models without magnetic fields do a good job of predicting the limb-darkening for WASP-18. This star is known to have abnormally low level of magnetic activity compared to similar stars of the same age, probably due to the influence of its massive, very short-period  planetary companion \citep[$M=10\,M_{\rm Jup}$, $P=0.94$\,d,][]{2014A&A...567A.128P, 2018AJ....155..113F}.

Magnetic activity will cause additional scatter in the light curve during the transit due to the variations in the mean flux level outside transit and occultation of active regions by the planet. This can be seen for the case of Kepler-17 in Fig.~\ref{fig:lcfit}. The quantity $r$ given in Table~\ref{tab:results2} quantifies this additional scatter. There is no clear correlation between the values of $\Delta h^{\prime}_1$ or $\Delta h^{\prime}_2$ and $r$, but there are only three stars with $r>1.2$ so it difficult to know how to interpret this result. 

An anonymous referee has suggested that the stellar atmosphere models used to compute the various limb-darkening tabulations in Table~\ref{tab:results2} may have been designed to agree with the measurements of the solar limb-darkening  by \citet{1994SoPh..153...91N}. This is not the case for the models by Kostogryz et al., which were tested  against the solar data but not calibrated using these data (A. Shapiro, priv. comm.), nor for the MARCS models used by \citet{2022RNAAS...6..248M} (B. Plez, priv. comm.). I was unable to find any mention of calibration against solar limb-darkening measurements in the papers that describe the models used for the limb-darkening tabulations in Table~\ref{tab:results2} \citep{1992IAUS..149..225K, 2013A&A...553A...6H, 2008A&A...491..633L}. 
Even if these models were ``tuned'' to agree with Neckel \& Labs' measurements, this would not lead to a good fit to the transit light curves of a solar-twin since this light curve would include the effect of the planet crossing spots and faculae on the stellar disc, features that were strictly excluded from the measurements reported in \citet{1994SoPh..153...91N}.

% Results table
\begin{table*}
\centering
\caption{Results from our light curve analysis. $b < 0.65$ }
	\label{tab:results2}

\begin{tabular}{@{}lllrrrrl} 
\hline		
  \multicolumn{1}{@{}l}{Model } & 
  \multicolumn{1}{l}{Source} & 
  \multicolumn{1}{c}{$\langle \Delta h^{\prime}_1\rangle$ } & 
  \multicolumn{1}{c}{$\sigma_{\rm ext,1}$ } & 
  \multicolumn{1}{c}{$\langle \Delta h^{\prime}_2\rangle$ } & 
  \multicolumn{1}{c}{$\sigma_{\rm ext,2}$ } & 
  \multicolumn{1}{c}{$N$ } 
  & Notes
   \\ 
\hline

\noalign{\smallskip}
\multicolumn{1}{@{}l}{\it Kepler} \\
Stagger-grid & \citet{2018A+A...616A..39M} &$ +0.007 \pm 0.002 $&$ 0.006 $&$ -0.007 \pm 0.004 $&  0.013 & 21 & \\
MPS-ATLAS    & \citet{2022arXiv220606641K} &$ +0.006 \pm 0.002 $&$ 0.004 $&$ -0.012 \pm 0.004 $&  0.012 & 24 & Set 1 \\
MPS-ATLAS    & \citet{2022arXiv220606641K} &$ +0.009 \pm 0.002 $&$ 0.005 $&$ -0.013 \pm 0.003 $&  0.013 & 24 & Set 2 \\
ATLAS        & \citet{2011A+A...529A..75C} &$ +0.006 \pm 0.002 $&$ 0.005 $&$ -0.013 \pm 0.004 $&  0.013 & 24 & \\
ATLAS        & \citet{2010A+A...510A..21S} &$ +0.003 \pm 0.001 $&$ 0.004 $&$ -0.013 \pm 0.003 $&  0.012 & 24 &  \\
sATLAS       & \citet{2013A+A...556A..86N} &$ +0.035 \pm 0.003 $&$ 0.010 $&$ -0.011 \pm 0.004 $&  0.014 & 24 & Mass $M = 1.1M_{\odot}$ \\
ATLAS        & \citet{2013A+A...556A..86N} &$ -0.022 \pm 0.004 $&$ 0.014 $&$ +0.036 \pm 0.004 $&  0.014 & 24 &  \\
PHOENIX-COND & \citet{2018A+A...618A..20C} &$ +0.010 \pm 0.003 $&$ 0.009 $&$ -0.003 \pm 0.004 $&  0.013 & 24 & Linear correction for [Fe/H]\\
MARCS        & \citet{2022RNAAS...6..248M} &$ +0.029 \pm 0.002 $&$ 0.009 $&$ -0.025 \pm 0.004 $&  0.015 & 23 & Linear correction for [Fe/H] \\
\noalign{\smallskip}
\multicolumn{1}{@{}l}{\it TESS} \\
Stagger-grid & \citet{2018A+A...616A..39M} &$ +0.006 \pm 0.004 $&$ 0.002 $ &$ -0.001 \pm 0.007 $& 0.005 &  6 & \\
MPS-ATLAS    & \citet{2022arXiv220606641K} &$ +0.004 \pm 0.003 $&$ 0.005 $ &$ -0.009 \pm 0.004 $& 0.002 & 10 & Set 1 \\
MPS-ATLAS    & \citet{2022arXiv220606641K} &$ +0.006 \pm 0.003 $&$ 0.005 $ &$ -0.010 \pm 0.004 $& 0.002 & 10 & Set 2 \\
ATLAS        & \citet{2017A+A...600A..30C} &$ +0.005 \pm 0.003 $&$ 0.005 $ &$ -0.011 \pm 0.004 $& 0.001 & 10 & Microturbulence $\xi = 2$\,km/s\\
PHOENIX-COND & \citet{2018A+A...618A..20C} &$ +0.011 \pm 0.004 $&$ 0.008 $ &$ -0.002 \pm 0.006 $& 0.005 & 10 & Linear correction for [Fe/H] \\
\noalign{\smallskip}\hline
\end{tabular}
\end{table*}

% EarthSunTransitSimulation figure
\begin{figure}
	\includegraphics[width=\columnwidth]{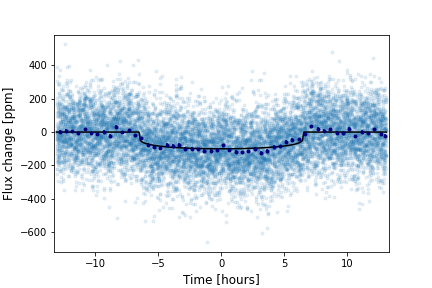}
   \caption{A simulated PLATO light curve of a bright Sun-like star with transits by an Earth-like planet. All three simulated transits are shown together as function of time relative to the closest time of mid-transit. Light-blue points are the simulated data at a cadence of 25 seconds. Dark-blue points show the average values in 30-minute bins. The model transit used to inject the transits into the simulated light curve is shown as a black line. }
    \label{fig:EarthSunTransitSimulation}
\end{figure}

\subsection{Implications for the PLATO mission}
It is beyond the scope of this study to explore these implications for the full range of known exoplanet systems that can now be studied with a wide variety of ground-based and space-based instrumentation. However, this study was motivated by the International Space Science Institute (ISSI) International Teams project  "Getting Ultra-Precise Planetary Radii with PLATO: The Impact of Limb Darkening and Stellar Activity on Transit Light Curves", so it is worthwhile to consider in the light of these results whether uncertainties in limb-darkening models are a significant obstacle to the primary aim of the PLATO mission -- to measure the radii of Earth-like planets in the habitable zones of Sun-like stars with a precision of 3\,per~cent \citep{2016AN....337..961R}. 

I used the PLATO solar-like light-curve simulator {\tt psls} version 1.5 \citep{2019A&A...624A.117S} to generate 1000 days of simulated data for a Sun-like star with an apparent magnitude $V=10$ assuming that the star is observed by all 24 cameras. Apart from the apparent magnitude of the star and the time span of the data, all other options were left at the values set in the example input configuration file {\tt psls.yaml} provided with the software. Simulated trends in the data were removed by dividing the simulated flux values by a smoothed version of the light curve using a Savitzky-Golay filter with a window width of 1 day.
 
For each trial in this Monte Carlo analysis, I selected a random time of mid-transit for the first transit and assumed that three transits 1 year apart are observed consecutively. The model transit is calculated assuming that the planet has the same radius and orbital period as the Earth, the transit impact parameter is $b=0$, the orbit is circular, and that the star has the same mass and radius as the Sun. The model transit was calculated using the methods described in Section~\ref{sec:simulations}. A typical simulated light curve is shown in Fig.~\ref{fig:EarthSunTransitSimulation}

To select a limb-darkening law for the analysis, I fitted the noiseless model transit light curve assuming either a quadratic limb-darkening law, a power-2 limb-darkening law, or the Claret 4-parameter limb-darkening law. I selected the power-2 law since it shows the lowest standard deviation of the residuals of these three limb-darkening laws (0.017\,ppm cf. 0.023\,ppm for the 4-parameter law and 0.106\,ppm for the quadratic law). The power-2 law also shows the smallest offset between the planet-star radius ratio derived from a least-squares fit to the noiseless transit light curve and the  planet-star radius ratio 
used to simulate the transit light curve ($-0.01$\,per~cent cf. 0.10\,per-cent for the 4-parameter law and 0.16\,per~cent for the quadratic law).

Least-squares fits to each of the simulated light curves were performed with Gaussian priors on $h_1^{\prime}$ and $h_2^{\prime}$ centred on the values determined from the limb-darkening profile used to simulate the model transit and with standard errors of 0.01 on $h_1^{\prime}$  and 0.02 $h_2^{\prime}$. These nominal uncertainties on $h_1^{\prime}$ and $h_2^{\prime}$ are based on the results in Table~\ref{tab:results2} assuming that an empirical correction is applied to the values of these parameters from one of the models that gives accurate predictions of their values, but with some uncertainty due to the observed scatter around the predicted values and the standard error in the zero-point correction. The free parameters in these least-squares fits are $k=R_{\rm pl}/R_{\star}$, $b$, $R_{\star}/a$, $h_1^{\prime}$ and $h_2^{\prime}$. I also used a Gaussian prior on the mean stellar density calculated from $R_{\star}/a$ and $P$ via Kepler's third law assuming that this value is known accurately with a precision of 1\,per~cent from asterosiesmology of the host star. These results were compared to the results of similar least-squares fits to the same simulated light curves with the limb-darkening fixed to the best-fit power-2 law to the actual limb-darkening profile.  

From 10\,000 Monte Carlo simulations we find that the standard error on the planet-star radius ratio with fixed limb-darkening is 2.579\,per~cent whereas with our nominal Gaussian priors on $h_1^{\prime}$ and $h_2^{\prime}$ the standard error on the planet-star radius ratio is 2.594\,per~cent. This shows that uncertainties on limb-darkening will add approximately (0.3\,per~cent)$^2$ to the variance in the measured planet-star radius ratio for Earth-like planets orbiting in the habitable zones of bright Sun-like stars, i.e. much smaller than the expected error on the measured planet-star ratio for Earth-like planets orbiting bright Sun-like stars.

\section{Conclusions}
\label{sec:conclusions}
 In this study, I have studied the limb-darkening information content in the {\it Kepler} and {\it TESS} light curves for a sample of solar-type stars with transiting hot-Jupiter companions. The limb-darkening information is quantified using the parameters $h^{\prime}_1 =  I_{\lambda}(\nicefrac{2}{3})$ and $h^{\prime}_2 =  h^{\prime}_1 - I_{\lambda}(\nicefrac{1}{3})$. These parameters are shown to be well-defined and subject to little or no systematic error. These observed values of $h^{\prime}_1$ and $h^{\prime}_2$ have been compared to the predictions from several grids of stellar atmosphere models. In general, the agreement between models and observations is very good. There is a small but significant offset $\Delta h^{\prime}_1 \approx 0.006$ between the observed and calculated values of $h^{\prime}_1$ which can be ascribed to the impact of the magnetic field on the atmospheric structure of these solar-type stars.

 Based on these results, I recommend that any of the following sources can be used to obtain reliable limb-darkening data for solar-type stars with low levels of magnetic activity -- \citet{2018A+A...616A..39M}, \citet{2022arXiv220606641K},  \citet{2011A+A...529A..75C}, \citet{2010A+A...510A..21S}. The performance of all these models is similar and none of them show significant trends in $\Delta h^{\prime}_1$ or $\Delta h^{\prime}_2$ with $T_{\rm eff}$. The ``Set 1'' models from \citet{2022arXiv220606641K} perform particularly well and cover a wide range of $T_{\rm eff}$, $\log g$ and [Fe/H] values, and with good sampling of this parameter space. However, it should be noted that all these models will show a small offset from the true limb-darkening profile because they do not account for the effects of the mean magnetic field on the star's atmospheric structure. This will introduce a small systematic error in the parameters obtained from the analysis of the light curve if the limb-darkening profile is fixed in the analysis. A better approach is to include some or all of the limb-darkening coefficients as free parameters in the analysis but with Gaussian priors on the values of $h^{\prime}_1$ and $h^{\prime}_2$ included in the fit. For $h^{\prime}_1$, the mean of the Gaussian prior should include a small correction based on the value of $\langle \Delta h^{\prime}_1\rangle$ for the model used from Table~\ref{tab:results2}. The standard error on the Gaussian prior for $h^{\prime}_1$ should account for the uncertainties in the values of $T_{\rm eff}$, $\log g$ and [Fe/H] used to estimate the limb-darkening coefficients, plus some additional error to account for the uncertainty in the correction, plus the scatter around the predicted values ($\sigma_{\rm ext, 1}$). The same approach can be used for the Gaussian prior on $h^{\prime}_2$ but an additional error $\approx \pm 0.005$ should be included to allow for the small systematic error in $h^{\prime}_2$ due to differences in the data pre-processing and data analysis.

 For example, according to the ``Set 1'' models from  \citet{2022arXiv220606641K}, a star with $T_{\rm eff} = 6000 \pm 100$\,K, $\log g = 4.0 \pm 0.05$  and ${\rm [Fe/H]} = 0.0 \pm 0.1 $
will have $h^{\prime}_1 = 0.874 \pm 0.004  $, $h^{\prime}_2 = 0.168 \pm 0.004$ in the {\it Kepler} band.  The empirical correction to the value of $h^{\prime}_1$ from Table~\ref{tab:results2} is $\langle \Delta h^{\prime}_1\rangle = +0.006 \pm 0.002 $ with a scatter $\sigma_{\rm ext, 1} =  0.004 $. The Gaussian prior to be applied for the analysis of the light curve would then be $h^{\prime}_1 = 0.880 \pm 0.006$. For  $h^{\prime}_2$, $\langle \Delta h^{\prime}_2\rangle = -0.012 \pm 0.004 $ with a scatter $\sigma_{\rm ext, 2} =  0.012 $. The Gaussian prior to be applied for the analysis of the light curve including an additional error of 0.005 to account for uncertainties due to data pre-processing and analysis method differences would then be $h^{\prime}_{2} = 0.156 \pm 0.014$.

 I have used a Monte Carlo simulation of PLATO light curves to show that limb-darkening will not be a significant contribution to the uncertainties in planet radii measured by the PLATO mission provided that the limb-darkening models are carefully selected and calibrated against observations of real stars using measurements similar to those presented here.
 
\section*{Acknowledgements}

This research was inspired by discussions with colleagues organised by the International Space Science Institute (ISSI) as part of the International Teams project 493 ``Getting Ultra-Precise Planetary Radii with PLATO: The Impact of Limb Darkening and Stellar Activity on Transit Light Curves'' (ISSI Team led by Szilárd Csizmadia).

This research was supported by UK Science and Technology Facilities Council (STFC) research grant number ST/M001040/1.

This research made use of Lightkurve, a Python package for Kepler and TESS data analysis \citep{2018ascl.soft12013L}.

This paper includes data collected by the Kepler and TESS missions obtained from the MAST data archive at the Space Telescope Science Institute (STScI). Funding for the Kepler mission is provided by the NASA Science Mission Directorate. Funding for the TESS mission is provided by the NASA Explorer Program. STScI is operated by the Association of Universities for Research in Astronomy, Inc., under NASA contract NAS 5–26555.

This research has made use of the VizieR catalogue access tool, CDS, Strasbourg, France (DOI : 10.26093/cds/vizier). 

I thank Nadiia Kostogryz and Alexander Shapiro for providing me with their tables of limb-darkening data from Kostogryz et al. (2022) prior to publication, and for very useful discussions related to their use of the limb-darkening data from Neckel and Labs (1994) to test the accuracy of the limb-darkening profiles from their stellar model atmospheres.

I thank David Sing for pointing me to the discussion in Kurucz (1992) of how the ATLAS stellar models used in Sing (2010) were calibrated against solar irradiance data.

I thank an anonynous referee for their comments on the manuscript that have helped to improve the paper.

%%%%%%%%%%%%%%%%%%%%%%%%%%%%%%%%%%%%%%%%%%%%%%%%%%
\section*{Data Availability}

The data underlying this article are available in the MAST data archive at the Space Telescope Science Institute (STScI) at \url{https://archive.stsci.edu} or from VizieR catalogue access tool hosted by the Centre de Données astronomiques de Strasbourg at \url{https://vizier.cds.unistra.fr/}



\bibliographystyle{mnras}
\bibliography{mybib} % if your bibtex file is called example.bib

%\begin{thebibliography}{99}
%\end{thebibliography}

%%%%%%%%%%%%%%%%%%%%%%%%%%%%%%%%%%%%%%%%%%%%%%%%%%

%%%%%%%%%%%%%%%%% APPENDICES %%%%%%%%%%%%%%%%%%%%%

%\appendix

%\section{Comparison to published limb-darkening laws}
%This appendix shows the results of comparing the observed values of $h^{\prime}_1$ and $h^{\prime}_2$ to various published limb-darkening laws as a function of the star's effective temperature


%%%%%%%%%%%%%%%%%%%%%%%%%%%%%%%%%%%%%%%%%%%%%%%%%%

% Don't change these lines
\bsp	% typesetting comment
\label{lastpage}
\end{document}